\documentclass[preprint,showpacs,preprintnumbers,amsmath,amssymb,prb,eqsecnum]{revtex4}
\usepackage[latin9]{inputenc}
\usepackage{amsmath}
\usepackage{graphicx}
\usepackage{esint}

\makeatletter

\providecommand{\tabularnewline}{\\}

\@ifundefined{textcolor}{}
{%
 \definecolor{BLACK}{gray}{0}
 \definecolor{WHITE}{gray}{1}
 \definecolor{RED}{rgb}{1,0,0}
 \definecolor{GREEN}{rgb}{0,1,0}
 \definecolor{BLUE}{rgb}{0,0,1}
 \definecolor{CYAN}{cmyk}{1,0,0,0}
 \definecolor{MAGENTA}{cmyk}{0,1,0,0}
 \definecolor{YELLOW}{cmyk}{0,0,1,0}
}

\usepackage{dcolumn}
\usepackage{bm}

\makeatother

\begin{document}

\title{Resonant x-ray scattering from chiral materials,\\
 $\alpha$-quartz and $\alpha$-berlinite}

\author{Jun-ichi Igarashi}

\affiliation{Faculty of Science, Ibaraki University, Mito, Ibaraki 310-8512, Japan}

\author{Manabu Takahashi}

\affiliation{Faculty of Engineering, Gunma University, Kiryu, Gunma 376-8515,
Japan}
\begin{abstract}
We study the resonant x-ray scattering at Si and Al K-edges
from chiral materials, $\alpha$-quartz and $\alpha$-berlinite. We
derive the general form of the scattering matrix for the dipole transition
by summing up the local scattering matrices which satisfy the symmetry
requirement. The oscillation term is obtained in the spectral intensity
as a function of azimuthal angle with an expression of possible phase
shift. We evaluate the parameters undetermined by the symmetry argument
alone on the basis of underlying electronic structures given by the
bond-orbital model. The spectra are calculated on forbidden spots
$\left(001\right)$, $\left(00\overline{1}\right)$, $\left(002\right)$,
and $\left(00\overline{2}\right)$ in circular polarizations without
adjustable parameter, reproducing well the experimental curves depending
on polarization, chirality, and scattering vector. Some discrepancies
remain in the phase shift in $\alpha$-quartz.
\end{abstract}


\maketitle

\section{\label{sect.1}Introduction}

The $\alpha$-quartz (SiO$_{2}$) and $\alpha$-berlinite (AlPO$_{4}$)
are known to have two crystal forms, the right-handed screw (space
group No.152, $P3_{1}21$) and the left-handed screw (No.154, $P3_{2}21$).
One is the mirror image of the other, thus the crystals are called
to have different chirality. The two forms have been distinguished
by using the optical activity since the discovery of Arago and Biot;\cite{Mason1982}
as linearly polarized light passes through the crystal, the direction
of polarization rotates about the beam axis oppositely according to
the chirality. Another method to distinguish chirality is the anomalous
x-ray scattering by which atomic scattering amplitudes become complex
numbers, \cite{Bijvoet1951,Vries1958} leading to the determination
of atomic positions for systems with different chirality. Note that
the conventional x-ray diffraction using Thomson scattering could
not distinguish chirality.

\begin{figure}
\includegraphics[width=8cm]{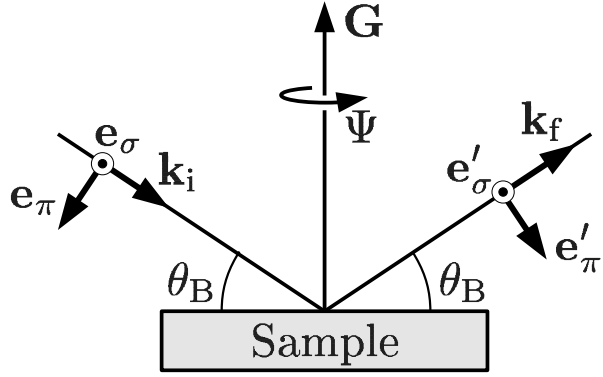}\caption{\label{fig.scat.geom} Scattering geometry. The wave vectors ${\bf k}_{i}$
and ${\bf k}_{f}$ are for the incident and scattered photons, respectively.
The scattering vector is defined by ${\bf G}={\bf k}_{f}-{\bf k}_{i}$.
The ${\bf e}_{\sigma}$, ${\bf e}_{\pi}$ are polarization vectors
for the incident photon, and ${\bf e}'_{\sigma}$ and ${\bf e}'_{\pi}$
are for the scattered photon. The ${\bf e}_{\sigma}$, ${\bf e}'_{\sigma}$
are perpendicular to the scattering plane. Three vectors $({\bf e}_{\sigma},{\bf e}_{\pi},{\bf k}_{i})$
and $({\bf e}'_{\sigma},{\bf e}'_{\pi},{\bf k}_{f})$ constitute a
right-handed coordinate frame, respectively. The sample is right-handedly
rotated around ${\bf G}$ by azimuthal angle $\Psi$. }
\end{figure}

Recently, another method has been attempted to distinguish chirality:
the resonant x-ray scattering (RXS) with circularly polarized beam.
\cite{Tanaka2008,Tanaka2010} The RXS process may be described at
the Si K-edge in $\alpha$-quartz and at the Al K-edge in $\alpha$-berlinite
as follows. The $1s$-core electron is prompted to unoccupied $p$-symmetric
states by absorbing photon [electric-dipole (E1) transition], and
subsequently the excited electron is recombined to the core-hole by
emitting photon (E1 transition). This will be called the E1-E1 process.
Figure \ref{fig.scat.geom} shows the scattering geometry, where the
sample is rotated by azimuthal angle $\Psi$ around the scattering
vector ${\bf G}={\bf k}_{f}-{\bf k}_{i}$. It is known that the tensor
character of the scattering matrix of RXS could give rise to the intensity
on the spots forbidden in Thomson scattering.\cite{Templeton1982,Dmitrienko1983,Dmitrienko2005}
Measuring the spectra on the forbidden spots ${\bf G}=\left(001\right)$
and $\left(00\overline{1}\right)$, Tanaka {\sl et al.}\cite{Tanaka2008,Tanaka2010}
have found characteristic patterns depending on chirality in the spectral
intensity by means of switching polarizations from the right-handed
one (RCP) to the left-handed one (LCP).

The purpose of this paper is to analyze such spectra from underlying
electronic structures. We start by deriving the scattering matrix
on the basis of symmetry requirement; the only assumption made is that
the total resonant scattering matrix is the sum of the local scattering matrix
\cite{Templeton1982,Dmitrienko1983,Dmitrienko2005} on each Si or
Al site. We introduce the local dipole-dipole correlation function
instead of the local scattering matrix in this procedure, since the
former quantity consisting of real numbers makes the expression transparent.
We obtain a general formula of scattering matrix depending on polarization,
chirality, and scattering vector. The scattering intensity contains
the oscillation term as a function of azimuthal angle, which has a
phase shift and the amplitude with chirality dependence different
from previous ones. \cite{Lovesey2008,Tanaka2010,Tanaka2010-2,Tanaka2012}

The general expression of scattering matrix involves parameters undetermined
by the symmetry argument alone. In this paper, we evaluate these parameters
from underlying electronic structures by exploiting a simple bond-orbital
model developed by Harrison.\cite{Harrison} This model is known to
work well on the ground-state properties in covalent-bonding systems;
it considers the strong coupling between the sp$^{3}$-hybrid on Si
atoms and the sp$^{1.24}$-hybrid on O atoms with four bonds per Si
atom in $\alpha$-quartz. In $\alpha$-berlinite, Si atoms are replaced
by Al and P atoms. In the ground state, the bonding states are occupied,
while the anti-bonding states are unoccupied. In the intermediate
state, one anti-bonding state is occupied at the core-hole site, on
which the core-hole potential works. The parameters in the scattering
matrix are evaluated by using the electronic structures thus determined.
The spectral intensities are calculated depending on polarization,
scattering vectors, and chirality as a function of azimuthal angle,
reproducing the experimental curves particularly well for $\alpha$-berlinite.
The present model calculation predicts the \emph{phase shift} of oscillation
to be $\pi$ for both $\alpha$-quartz and $\alpha$-berlinite. The
experimental curves indicate the phase shift $\pi$ in $\alpha$-berlinite
and its deviation from $\pi$ in $\alpha$-quartz. \cite{Tanaka2008,Tanaka2010}
Since both materials are expected to have similar electronic structures,
such a difference is puzzling to us. The possible origin for the deviation
will be discussed in the last section.

The present paper is organized as follows. In Sec. \ref{sect.2},
the crystal structures of $\alpha$-quartz and $\alpha$-berlinite
are briefly described. In Sec. \ref{sect.3}, the scattering matrix
is formulated from the symmetry requirement. In Sec. \ref{sect.4},
the RXS intensity as a function of azimuthal angle is formulated for
the incident x-ray beam specified by the Stokes parameters. In Sec.
\ref{sect.5}, the bond-orbital model is introduced to evaluate the
electronic structures as well as the RXS intensity. In Sec. \ref{sect.6},
the calculated results are discussed in comparison with experiments.
Section \ref{sect.7} is devoted to the concluding remarks.

\section{\label{sect.2}Crystal structure}

\subsection{$\alpha$-quartz}

\begin{figure}
\includegraphics[width=10cm]{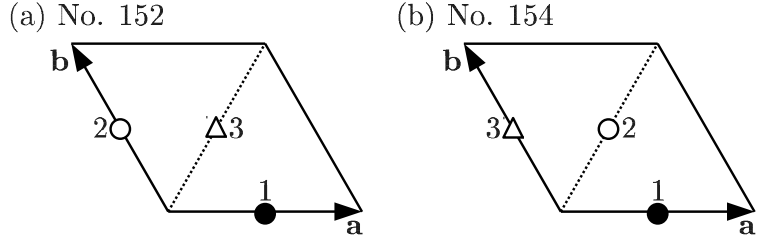}\caption{\label{fig.crystal} Si atoms projected onto the $a$-$b$ plane in
the unit cell of $\alpha$-quartz. Atoms labeled by 1, 2, 3 are located
at $(u,0,0)$, $(0,u,\frac{1}{3}c)$, $(1-u,1-u,\frac{2}{3}c)$ for
No.152, and at $(u,0,0)$, $(1-u,1-u,\frac{1}{3}c)$, $(0,u,\frac{2}{3}c)$,
respectively. $u=0.47$. Si atoms are surrounded by O atoms forming
a tetrahedron. }
\end{figure}

The crystal structure of $\alpha$-quartz is hexagonal with three
Si atoms per unit cell with $a=b=4.91$\AA{}, and $c=5.40$\AA{}.
It has chirality described by two different space group $P3_{1}21$
(No.152), the right-handed screw, and $P3_{2}21$ (No.154), the left-handed
screw. As shown in Fig.~\ref{fig.crystal}, Si atoms sit at the positions
$(u,0,0)$, $(0,u,\frac{1}{3})$, $(1-u,1-u,\frac{2}{3})$ for No.152,
while at the positions $(u,0,0)$, $(1-u,1-u,\frac{1}{3})$, $(0,u,\frac{2}{3})$
for No.154, where $u=0.47$. One O atom sits between each nearest-neighboring
Si-Si pair. Each Si atom is surrounded by a tetrahedron
of O atoms. We introduce the Cartesian frame where $x$ and $z$ axes
are parallel to the crystal $a$ and $c$ axes and the origin is set
at the center of the Si atom labeled as 1 in Fig. \ref{fig.crystal}.
Then, the coordinates of O atoms of the tetrahedron are given by $(0.93,\mp0.59,-1.13)$,
$(-0.96,1.13,\mp0.64)$, $(-0.96,-1.13,\pm0.64)$, $(0.93,\pm0.59,1.13)$
in units of \AA{}, where the upper and lower signs correspond to No.152
and No.154, respectively. \cite{Page1976} Note that there exists
the symmetry of two-fold rotation around the $a$-axis for both No.152
and No.154, and that the crystal of No.154 is the mirror image of
No.152 with respect to the $a$-$b$ plane. The settings correspond
to the $r\left(+\right)$ and $z\left(-\right)$ settings for the
crystals of No. 152 and No.154, respectively, in which the twofold
axis develops $\left(+\right)$ charge at its positive end on crystal
extension along the axis. Other tetrahedrons are given by rotating
$\frac{2\pi}{3}$ ($-\frac{2\pi}{3}$) and $\frac{4\pi}{3}$ ($-\frac{4\pi}{3}$)
around the $c$ axis with translating $\frac{1}{3}c$, $\frac{2}{3}c$
along the $c$ axis for No.152 (No.154).

\subsection{$\alpha$-berlinite}

The crystal structure is close to the $\alpha$-quartz structure,
where Si atoms are replaced by Al and P atoms alternatively along
the $c$-axis. Therefore, the unit cell is doubled along the $c$-axis;
there are three Al atoms and three P atoms per unit cell with $a=b=4.94$\AA{},
$c=10.95$\AA{}. More precisely, for No.152, Al atoms sit at $(u,0,0)$,
$(0,u,\frac{1}{3})$, $(1-u,1-u,\frac{2}{3})$, and P atoms at $(1-u',1-u',\frac{1}{6})$,
$(u',0,\frac{1}{2})$, $(0,u',\frac{5}{6})$, with $u\approx u'=0.47$,
while for No.154, Al atoms sit at $(u,0,0)$, $(1-u,1-u,\frac{1}{3})$,
$(0,u,\frac{2}{3})$, and P atoms at $(0,u',\frac{1}{6})$, $(u',0,\frac{1}{2})$,
$(1-u',1-u',\frac{5}{6})$. In the Cartesian frame where the $x$ and
$z$ axes are along the $a$ and $c$ axes, respectively, and its
origin is the center of the Al site at $(u,0,0)$, the coordinates of
the O atoms of the tetrahedron are given by $(1.01,\pm0.67,-1.25)$, $(-0.95,1.27,\pm0.69)$,
$(-0.95,-1.27,\mp0.69)$, $(1.01,\mp0.67,1.25)$ in units of \AA{},
where the upper (lower) sign corresponds to No.152 (No.154).\cite{Muraoka1997}
Note that the tetrahedron surrounding Al for No.154 is similar to
that for No.152 in $\alpha$-quartz. Other tetrahedrons surrounding
Al are given by rotating $\frac{2\pi}{3}$ ($-\frac{2\pi}{3}$), $\frac{4\pi}{3}$
($-\frac{4\pi}{3}$) around the $c$-axis with translating $\frac{1}{3}c$,
$\frac{2}{3}c$ along the $c$-axis for No.152 (No.154), according
to the screw symmetry.

\section{\label{sect.3}Scattering matrix}

Let the incident and scattered photon polarizations be specified as
$x_{\beta}$ and $x_{\alpha}$ in Cartesian frame where the $x_{1}$($\equiv x$)
and $x_{3}$($\equiv z$) axes are along the $a$ and $c$ axes. The
scattering matrix at site $j$, $\hat{M}(j;\omega)$, may be expressed
as 
\begin{equation}
[\hat{M}(j;\omega)]_{\alpha,\beta}=\frac{\langle g|\hat{x}_{\alpha}|n\rangle\langle n|\hat{x}_{\beta}|g\rangle}{\omega+\epsilon_{g}-\epsilon_{n}+i\Gamma},
\end{equation}
where the dipole operator $\hat{x}_{\alpha}$ is measured from the
center of site $j$. Ket $|g\rangle$ represents the ground state
with energy $\epsilon_{g}$, and $|n\rangle$ represents the intermediate
state with energy $\epsilon_{n}$ (including the core-hole energy).
The $\Gamma$ represents the life-time broadening width of the $1s$-core
hole.

The total resonant scattering matrix could be well approximated by the sum
of the local amplitudes at Si sites or at Al sites, since the $1s$
state is localized at each Si site. The scattering geometry is shown
in Fig.~\ref{fig.scat.geom}. For the incident and scattered wave
vectors ${\bf k}_{i}$ and ${\bf k}_{f}$, the total resonant scattering matrix
may be expressed as 
\begin{equation}
[\hat{M}({\bf G},\omega)]_{\alpha,\beta}=\sum_{j}[\hat{M}(j;\omega)]_{\alpha,\beta}\exp(-i{\bf G\cdot r}_{j}).
\end{equation}
where $j$ runs over Si sites for Si $K$-edge (quartz) and over Al
sites for Al $K$-edge (berlinite). The scattering vector is defined
by ${\bf G}={\bf k}_{f}-{\bf k}_{i}$.

To analyze the symmetry of the scattering matrix, it may be convenient
to introduce the local dipole-dipole correlation function at the core-hole
site $j$, which is defined by 
\begin{equation}
[\hat{\rho}(j;\epsilon)]_{\alpha,\beta}=\sum_{n}\langle g|\hat{x}_{\alpha}|n\rangle\langle n|\hat{x}_{\beta}|g\rangle\delta(\epsilon+\epsilon_{g}-\epsilon_{n}),
\end{equation}
and the total dipole-dipole correlation function associated with ${\bf G}$,
which is defined by 
\begin{equation}
\hat{\rho}({\bf G};\epsilon)=\sum_{j}\hat{\rho}(j;\epsilon)\exp(-i{\bf G}\cdot{\bf r}_{j}).
\end{equation}
Using the latter quantity, we may express the total resonant scattering matrix
as 
\begin{equation}
\hat{M}({\bf G};\omega)=\int\frac{\hat{\rho}({\bf G};\epsilon)}{\omega-\epsilon+i\Gamma}{\rm d}\epsilon.\label{eq.total}
\end{equation}

Let the local dipole-dipole correlation function at site $(u,0,0)$
be $\hat{\rho}_{0}^{(\pm)}$ with $+$ and $-$ signs corresponding
to No.152 and No.154. It should take the following matrix form according
to the local symmetry: 
\begin{equation}
\hat{\rho}_{0}^{(\pm)}(\epsilon)=\left(\begin{array}{ccc}
a(\epsilon) & 0 & 0\\
0 & b(\epsilon) & \pm d(\epsilon)\\
0 & \pm d(\epsilon) & c(\epsilon)
\end{array}\right),\label{eq.local.density}
\end{equation}
where $a(\epsilon)$, $b(\epsilon)$, $c(\epsilon)$, $d(\epsilon)$
are real functions. The presence of the off-diagonal elements is due
to the lack of the inversion symmetry around the Si atom in $\alpha$-quartz
or the Al atom in $\alpha$-berlinite. The zero components are originated
from the symmetry of the two-fold rotation around the $a$-axis. The
$\pm$ signs are originated from the mirror-image relation with respect
to the $a$-$b$ plane between No.152 and No.154, respectively.

Now we define the local correlation function rotated by $\pm2\pi/3$
around the $c$-axis: 
\begin{eqnarray}
\hat{\rho}_{1}^{(\pm)}(\epsilon) & = & \hat{R}(2\pi/3)\hat{\rho}_{0}^{(\pm)}(\epsilon)\hat{R}^{-1}(2\pi/3),\\
\hat{\rho}_{-1}^{(\pm)}(\epsilon) & = & \hat{R}(-2\pi/3)\hat{\rho}_{0}^{(\pm)}(\epsilon)\hat{R}^{-1}(-2\pi/3),
\end{eqnarray}
where rotation matrix is defined by 
\begin{equation}
\hat{R}(\theta)=\left(\begin{array}{rrr}
\cos\theta & -\sin\theta & 0\\
\sin\theta & \cos\theta & 0\\
0 & 0 & 1
\end{array}\right).
\end{equation}
For the crystal of No.152, the local dipole-dipole correlation function
at $(0,u,\frac{1}{3}c)$ and at $(1-u,1-u,\frac{2}{3}c)$ are given
by $\hat{\rho}_{1}^{(+)}(\epsilon)$ and $\hat{\rho}_{-1}^{(+)}(\epsilon)$,
respectively. On the other hand, for the crystal of No.154, the local
dipole-dipole correlation function at $(1-u,1-u,\frac{1}{3}c)$ and
at $(0,u,\frac{2}{3}c)$ are given by $\hat{\rho}_{-1}^{(-)}(\epsilon)$
and $\hat{\rho}_{1}^{(-)}(\epsilon)$, respectively.

\subsubsection{${\bf G}=(001)$}

Summing up the local correlation function defined above with weight
$\exp(-i{\bf G}\cdot{\bf r}_{j})$, we obtain the total dipole-dipole
correlation functions for No.152 and No.154 as 
\begin{align}
\hat{\rho}((001);\epsilon)_{152} & =\hat{\rho}_{0}^{(+)}(\epsilon)+\hat{\rho}_{1}^{(+)}(\epsilon)\exp(-i2\pi/3)+\hat{\rho}_{-1}^{(+)}(\epsilon)\exp(i2\pi/3)\nonumber \\
 & =\frac{3}{4}\left(\begin{array}{rrr}
[a(\epsilon)-b(\epsilon)] & i[a(\epsilon)-b(\epsilon)] & 2id(\epsilon)\\
i[a(\epsilon)-b(\epsilon)] & -[a(\epsilon)-b(\epsilon)] & 2d(\epsilon)\\
2id(\epsilon) & 2d(\epsilon) & 0
\end{array}\right),\label{eq.rho152}\\
\hat{\rho}((001);\epsilon)_{154} & =\hat{\rho}_{0}^{(-)}(\epsilon)+\hat{\rho}_{1}^{(-)}(\epsilon)\exp(i2\pi/3)+\hat{\rho}_{-1}^{(-)}(\epsilon)\exp(-i2\pi/3)\nonumber \\
 & =\frac{3}{4}\left(\begin{array}{rrr}
[a(\epsilon)-b(\epsilon)] & -i[a(\epsilon)-b(\epsilon)] & 2id(\epsilon)\\
-i[a(\epsilon)-b(\epsilon)] & -(a(\epsilon)-b(\epsilon)) & -2d(\epsilon)\\
2id(\epsilon) & -2d(\epsilon) & 0
\end{array}\right).\label{eq.rho154}
\end{align}
Note that there exist only two independent components $a(\epsilon)-b(\epsilon)$
and $d(\epsilon)$ in the matrix. Hence, from Eq. (\ref{eq.total}),
the total resonant scattering matrix is written as 
\begin{equation}
\hat{M}((001);\omega)=\left(\begin{array}{rrr}
A(\omega) & \pm iA(\omega) & iB(\omega)\\
\pm iA(\omega) & -A(\omega) & \pm B(\omega)\\
iB(\omega) & \pm B(\omega) & 0
\end{array}\right),\label{eq.M001}
\end{equation}
with 
\begin{eqnarray}
A(\omega) & = & \frac{3}{4}\int\frac{a(\epsilon)-b(\epsilon)}{\omega-\epsilon+i\Gamma}{\rm d}\epsilon,\label{eq.A}\\
B(\omega) & = & \frac{3}{2}\int\frac{d(\epsilon)}{\omega-\epsilon+i\Gamma}{\rm d}\epsilon,\label{eq.B}
\end{eqnarray}
where the upper and lower signs correspond to No.152 and No.154, respectively.
Note that $A(\omega)$ and $B(\omega)$ are complex numbers because
of the presence of $\Gamma$.

\subsubsection{${\bf G}=(00\overline{1})$}

The total dipole-dipole correlation functions for No.152 and No.154
are given by 
\begin{eqnarray}
\hat{\rho}((00\overline{1});\epsilon)_{152} & = & \hat{\rho}_{0}^{(+)}(\epsilon)+\hat{\rho}_{1}^{(+)}(\epsilon)\exp(i2\pi/3)+\hat{\rho}_{-1}^{(+)}(\epsilon)\exp(-i2\pi/3),\\
\hat{\rho}((00\overline{1});\epsilon)_{154} & = & \hat{\rho}_{0}^{(-)}(\epsilon)+\hat{\rho}_{1}^{(-)}(\epsilon)\exp(-i2\pi/3)+\hat{\rho}_{-1}^{(-)}(\epsilon)\exp(+i2\pi/3).
\end{eqnarray}
Since $a(\epsilon)$, $b(\epsilon)$, $c(\epsilon)$, $d(\epsilon)$
in $\rho_{0}^{(\pm)}(\epsilon)$ and $\rho_{\pm1}^{(\pm)}(\epsilon)$
are real functions, we notice that the total resonant scattering amplitude
is obtained from $M((001);\omega)$ by replacing $\pm i$ by $\mp i$.
The result is 
\begin{equation}
\hat{M}((00\overline{1});\omega)=\left(\begin{array}{rrr}
A(\omega) & \mp iA(\omega) & -iB(\omega)\\
\mp iA(\omega) & -A(\omega) & \pm B(\omega)\\
-iB(\omega) & \pm B(\omega) & 0
\end{array}\right),\label{eq.M00-1}
\end{equation}
where the upper and lower signs correspond to No.152 and No.154, respectively.

\subsubsection{${\bf G}=(002)$ and $(00\overline{2})$}

Now that the phase factors $\exp(\pm i4\pi/3)$ are equivalent to
$\exp(\mp i2\pi/3)$, we immediately obtain the following relation:
\begin{eqnarray}
\hat{M}((002);\omega) & = & \hat{M}((00\overline{1});\omega),\label{eq.M002}\\
\hat{M}((00\overline{2});\omega) & = & \hat{M}((001);\omega).\label{eq.M00-2}
\end{eqnarray}

\section{\label{sect.4}Polarization analysis with rotating crystal}

Specifying the polarization vectors by ${\bf e}_{\sigma}$ (${\bf e}'_{\sigma}$)
and ${\bf e}_{\pi}$ (${\bf e}'_{\pi}$) for incident (scattered)
photon, we formally write the component of the total resonant scattering matrix
as 
\begin{align}
\hat{M}({\bf G};\omega)_{\sigma^{\prime}\sigma} & ={\bf e}_{\sigma}^{\prime\dagger}\cdot\hat{M}({\bf G};\omega)\cdot{\bf e}_{\sigma},\\
\hat{M}({\bf G};\omega)_{\pi^{\prime}\sigma} & ={\bf e}_{\pi}^{\prime\dagger}\cdot\hat{M}({\bf G};\omega)\cdot{\bf e}_{\sigma},
\end{align}
and so on. We will evaluate these values with rotating crystal in
the scattering geometry described below.

\subsection{${\bf G}=(001)$ and $(002)$}

\begin{figure}
\includegraphics[width=8cm]{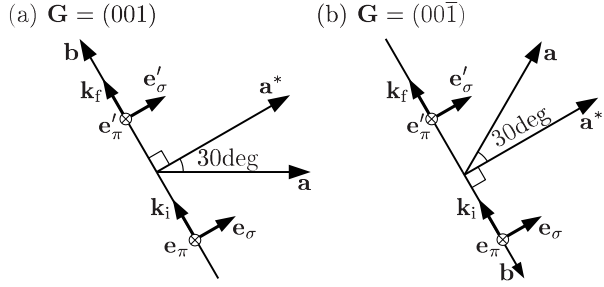}\caption{\label{fig.topview} Top view of the scattering geometry at the azimuthal
angle $\Psi=0$. (a)${\bf G}=(001)$: view looking down along the
$c$ axis from the top of the axis. (b)${\bf G}=(00\overline{1})$:
view looking up along the $c$ axis from the bottom of the axis. Vectors
${\bf a}$ and ${\bf b}$ are translational vectors along the $a$
and $b$ axes, respectively. Vector ${\bf a}^{*}$ is the reciprocal
lattice vector conjugate to ${\bf a}$. }
\end{figure}

We rotate the crystal right-handedly around ${\bf G}$ with azimuthal
angle $\Psi$. Following the experimental setup by Tanaka {\sl et al.},
\cite{Tanaka2008,Tanaka2010-2} we define the origin of $\Psi$ such
that the scattering plane contains the $b$ axis, that is, it is perpendicular
to the reciprocal lattice vector ${\bf a}^{*}$ conjugate to the translational
vector ${\bf a}$ along the $a$ axis, as shown in Fig.~\ref{fig.topview}(a).
Note that the rotation of the crystal indicates that the scattering
plane is inversely rotated with respect to the crystal.

The polarization vectors, which are represented in the Cartesian frame
with $x$ and $z$ axes along the crystal $a$ and $c$ axes, are
given by 
\begin{eqnarray}
{\bf e}_{\sigma} & = & {\bf e}'_{\sigma}=(-\sin\bar{\Psi},-\cos\bar{\Psi},0),\\
{\bf e}_{\pi} & = & (-\sin\theta_{{\rm B}}\cos\bar{\Psi},\sin\theta_{{\rm B}}\sin\bar{\Psi},-\cos\theta_{{\rm B}}),\\
{\bf e}'_{\pi} & = & (\sin\theta_{{\rm B}}\cos\bar{\Psi},-\sin\theta_{{\rm B}}\sin\bar{\Psi},-\cos\theta_{{\rm B}}),
\end{eqnarray}
where $\bar{\Psi}=\Psi-2\pi/3$, and $\theta_{{\rm B}}$ is the Bragg
angle of the scattering. Using Eq.~(\ref{eq.M001}), we obtain 
\begin{eqnarray}
\hat{M}((001);\omega)_{\sigma'\sigma} & = & -A{\rm e}^{\mp i2\bar{\Psi}},\label{eq.Mss}\\
\hat{M}((001);\omega)_{\pi'\sigma} & = & \mp iA\sin\theta_{{\rm B}}{\rm e}^{\mp i2\bar{\Psi}}\pm B\cos\theta_{{\rm B}}{\rm e}^{\pm i\bar{\Psi}},\\
\hat{M}((001);\omega)_{\sigma'\pi} & = & \pm iA\sin\theta_{{\rm B}}{\rm e}^{\mp i2\bar{\Psi}}\pm B\cos\theta_{{\rm B}}{\rm e}^{\pm i\bar{\Psi}},\\
\hat{M}((001);\omega)_{\pi'\pi} & = & -A\sin^{2}\theta_{{\rm B}}{\rm e}^{\mp i2\bar{\Psi}},\label{eq.Mpipi}
\end{eqnarray}
where the upper (lower) sign corresponds to the crystal of No.152
(No.154). We have abbreviated $A(\omega)$ and $B(\omega)$ by $A$
and $B$, respectively. For ${\bf G}=(002)$, the corresponding components
of scattering matrix are given by replacing $\pm i$ by $\mp i$ in
Eqs.~(\ref{eq.Mss})-(\ref{eq.Mpipi}).

\subsubsection{Linear polarization}

For the scattering channels with $\sigma\to\sigma'$ and $\pi\to\pi'$,
the scattering intensities are the same for both No.152 and No.154,
and are constant as a function of azimuthal angle. On the other hand,
for the scattering channels with $\sigma\to\pi'$ and $\pi\to\sigma'$,
the scattering intensities oscillate as a function of azimuthal angle.

For ${\bf G}=(001)$, the scattering intensities in the $\sigma\to\sigma'$
and $\sigma\to\pi'$ channels are given by 
\begin{eqnarray}
I((001);\omega)_{\sigma'\sigma} & = & |A|^{2},\label{eq.ss}\\
I((001);\omega)_{\pi'\sigma} & = & |A|^{2}\sin^{2}\theta_{{\rm B}}+|B|^{2}\cos^{2}\theta_{{\rm B}}\mp|AB|\sin2\theta_{{\rm B}}\sin(3\Psi\mp\delta),\label{eq.spi}
\end{eqnarray}
where the upper sign (lower) sign corresponds to the crystal of No.152
(No.154), and $B^{*}A$ is replaced by $|AB|{\rm e}^{i\delta}$. The
phase $\delta$ could take any value in principle.

For ${\bf G}=(002)$, the corresponding intensities are given by Eqs.~(\ref{eq.ss})
and (\ref{eq.spi}) with replacing $\sin(3\Psi\mp\delta)$ by $\sin(3\Psi\pm\delta)$.
Since $3\bar{\Psi}=3\Psi-2\pi$, we could safely replace $\bar{\Psi}$
by $\Psi$. The constant term as well as the amplitudes of the oscillation
are the same for both No.152 and No.154. Therefore, the knowledge
of the phase shift $\delta$ of the oscillation is necessary in order
to distinguish the chirality.

\subsubsection{Circular polarization of incident beam}

We consider the case where the incident beam is circularly polarized.
To include the partial polarization, we introduce the polarization
density matrix,\cite{Landau} which is represented on the bases ${\bf e}_{\sigma}$
and ${\bf e}_{\pi}$: 
\begin{equation}
\hat{P}=\frac{1}{2}\left(\begin{array}{cc}
1+P_{3} & P_{1}-iP_{2}\\
P_{1}+iP_{2} & 1-P_{3}
\end{array}\right),
\end{equation}
where $P_{1}$, $P_{2}$, $P_{3}$ are the Stokes parameters. All
three parameters take values between $-1$ and $+1$. In the unpolarized
state, $P_{1}=P_{2}=P_{3}=0$; for a completely polarized photon,
$P_{1}^{2}+P_{2}^{2}+P_{3}^{2}=1$; $P_{2}=+1(-1)$ for the right(left)-handed
circular polarization.

Without analyzing the polarization of scattered x-ray, the scattering
intensity may be expressed in the following form: 
\begin{eqnarray}
I({\bf G};\omega) & = & \sum_{n}\sum_{n_{1}}\sum_{n_{2}}\hat{M}_{nn_{1}}\hat{M}_{nn_{2}}\hat{P}_{n_{2}n_{1}},\nonumber \\
 & = & U_{1}P_{1}+U_{2}P_{2}+U_{3}^{+}\frac{1}{2}(1+P_{3})+U_{3}^{-}\frac{1}{2}(1-P_{3}).\label{eq.azim}
\end{eqnarray}

For ${\bf G}=(001)$, the coefficients are given by 
\begin{eqnarray}
U_{1} & = & \mp|AB|\cos\theta_{{\rm B}}(1+\sin^{2}\theta_{{\rm B}})\cos(3\Psi\mp\delta),\label{eq.U1}\\
U_{2} & = & \pm|A|^{2}\sin\theta_{{\rm B}}(1+\sin^{2}\theta_{{\rm B}})+|AB|\cos^{3}\theta_{{\rm B}}\sin(3\Psi\mp\delta),\label{eq.U2}\\
U_{3}^{+} & = & |A|^{2}(1+\sin^{2}\theta_{{\rm B}})+|B|^{2}\cos^{2}\theta_{{\rm B}}\mp|AB|\sin2\theta_{{\rm B}}\sin(3\Psi\mp\delta),\label{eq.U3+}\\
U_{3}^{-} & = & |A|^{2}\sin^{2}\theta_{{\rm B}}(1+\sin^{2}\theta_{{\rm B}})+|B|^{2}\cos^{2}\theta_{{\rm B}}\pm|AB|\sin2\theta_{{\rm B}}\sin(3\Psi\mp\delta).\label{eq.U3-}
\end{eqnarray}
Thereby the scattering intensity with $P_{1}=0$ is expressed as 
\begin{equation}
I((001);\omega)=I_{0}((001);\omega)+I_{1}((001);\omega)\sin(3\Psi\mp\delta),\label{eq.I001}
\end{equation}
with 
\begin{eqnarray}
I_{0}((001);\omega) & = & |B]^{2}\cos^{2}\theta_{{\rm B}}+|A|^{2}\frac{1+\sin^{2}\theta_{{\rm B}}}{2}\nonumber \\
 & \times & (1+\sin^{2}\theta_{{\rm B}}\pm2P_{2}\sin\theta_{{\rm B}}+P_{3}\cos^{2}\theta_{{\rm B}}),\\
I_{1}((001);\omega) & = & |AB|\cos\theta_{{\rm B}}(P_{2}\cos^{2}\theta_{{\rm B}}\mp2P_{3}\sin\theta_{{\rm B}}).\label{eq.I001_1}
\end{eqnarray}

For ${\bf G}=(002)$, the corresponding coefficients are given by
the same forms of Eqs.~(\ref{eq.U1})-(\ref{eq.U3-}) with replacing
$3\Psi\mp\delta$ by $3\Psi\pm\delta$ and with reversing the signs
in Eq.~(\ref{eq.U2}). Hence the scattering intensity with $P_{1}=0$
is expressed as 
\begin{equation}
I((002);\omega)=I_{0}((002);\omega)+I_{1}((002);\omega)\sin(3\Psi\pm\delta),\label{eq.I002}
\end{equation}
with 
\begin{eqnarray}
I_{0}((002);\omega) & = & |B]^{2}\cos^{2}\theta_{{\rm B}}+|A|^{2}\frac{1+\sin^{2}\theta_{{\rm B}}}{2}\nonumber \\
 & \times & (1+\sin^{2}\theta_{{\rm B}}\mp2P_{2}\sin\theta_{{\rm B}}+P_{3}\cos^{2}\theta_{{\rm B}}),\\
I_{1}((002);\omega) & = & |AB|\cos\theta_{{\rm B}}(-P_{2}\cos^{2}\theta_{{\rm B}}\mp2P_{3}\sin\theta_{{\rm B}}).\label{eq.I002_1}
\end{eqnarray}

\subsection{${\bf G}=(00\overline{1})$ and ${\bf G}=(00\overline{2})$}

Following the experimental setup by Tanaka {\sl et al.},\cite{Tanaka2008,Tanaka2010}
we rotate the crystal by angle $\pi$ around the reciprocal lattice
vector ${\bf a}^{*}$ in order to use the reverse side of the crystal
for the scattering, as shown in Fig.~\ref{fig.topview}(b). The other
scattering conditions are kept the same as those for ${\bf G}=(001)$
and $(002)$. Since the direction of ${\bf G}$ is now opposite to
$(001)$, the right-handed rotation of crystal around the direction
of ${\bf G}$ means the left-handed rotation around the $c$ axis.
The polarization vectors, which are represented in the Cartesian frame
with the $x$ and $z$ axes along the crystal $a$ and $c$ axes,
are given by 
\begin{eqnarray}
{\bf e}_{\sigma} & = & {\bf e}'_{\sigma}=(\sin\bar{\Psi},-\cos\bar{\Psi},0),\\
{\bf e}_{\pi} & = & (\sin\theta_{{\rm B}}\cos\bar{\Psi},\sin\theta_{{\rm B}}\sin\bar{\Psi},\cos\theta_{{\rm B}}),\\
{\bf e}'_{\pi} & = & (-\sin\theta_{{\rm B}}\cos\bar{\Psi},-\sin\theta_{{\rm B}}\sin\bar{\Psi},\cos\theta_{{\rm B}}),
\end{eqnarray}
where $\bar{\Psi}=\Psi+2\pi/3$. Thereby, from Eq.~(\ref{eq.M00-1}),
we obtain 
\begin{eqnarray}
\hat{M}((00\overline{1});\omega)_{\sigma'\sigma} & = & -A{\rm e}^{\mp i2\bar{\Psi}},\label{eq.Mss-1}\\
\hat{M}((00\overline{1});\omega)_{\pi'\sigma} & = & \mp iA\sin\theta_{{\rm B}}{\rm e}^{\mp i2\bar{\Psi}}\mp B\cos\theta_{{\rm B}}{\rm e}^{\pm i\bar{\Psi}},\\
\hat{M}((00\overline{1});\omega)_{\sigma'\pi} & = & \pm iA\sin\theta_{{\rm B}}{\rm e}^{\mp i2\bar{\Psi}}\mp B\cos\theta_{{\rm B}}{\rm e}^{\pm i\bar{\Psi}},\\
\hat{M}((00\overline{1});\omega)_{\pi'\pi} & = & -A\sin^{2}\theta_{{\rm B}}{\rm e}^{\mp i2\bar{\Psi}}.\label{eq.Mpipi-1}
\end{eqnarray}
These equations are nothing but Eqs.~(\ref{eq.Mss})-(\ref{eq.Mpipi})
with reversing signs in front of $B$. For ${\bf G}=(00\overline{2})$,
$\hat{M}((00\overline{2});\omega)_{\sigma'\sigma}\cdots$ are given
by the same forms of Eqs.~(\ref{eq.Mss-1})-(\ref{eq.Mpipi-1}) with
replacing $\pm i$ by $\mp i$. With these expressions, the following
results straightforwardly come out.

\subsubsection{Linear polarization}

We obtain the scattering intensity, 
\begin{equation}
I((00\overline{1});\omega)_{\pi'\sigma}=|A|^{2}\sin^{2}\theta_{{\rm B}}+|B|^{2}\cos^{2}\theta_{{\rm B}}\pm|AB|\sin2\theta_{{\rm B}}\sin(3\Psi\mp\delta).\label{eq.spi-1}
\end{equation}
which is the same as Eq.~(\ref{eq.spi}) except for reversing the
sign for the term proportional to $\sin({3\Psi\mp\delta})$. The scattering
intensity for ${\bf G}=(00\overline{2})$ is given by Eq.~(\ref{eq.spi-1})
with replacing $\sin(3\Psi\mp\delta)$ by $\sin(3\Psi\pm\delta)$.

\subsubsection{Circular polarization}

The $U_{1}$, $U_{2}$, $U_{3}^{+}$ and $U_{3}^{-}$ for ${\bf G}=(00\overline{1})$
are given by the same forms as Eqs.~(\ref{eq.U1})-(\ref{eq.U3-})
with reversing the signs of the terms proportional to $\cos(3\Psi\mp\delta)$
and $\sin(3\Psi\mp\delta)$. Thereby, the scattering intensity with
$P_{1}=0$ is expressed as 
\begin{equation}
I((00\overline{1});\omega)=I_{0}((00\overline{1});\omega)+I_{1}((00\overline{1});\omega)\sin(3\Psi\mp\delta),\label{eq.I00-1}
\end{equation}
with 
\begin{eqnarray}
I_{0}((00\overline{1});\omega) & = & |B]^{2}\cos^{2}\theta_{{\rm B}}+|A|^{2}\frac{1+\sin^{2}\theta_{{\rm B}}}{2}\nonumber \\
 & \times & (1+\sin^{2}\theta_{{\rm B}}\pm2P_{2}\sin\theta_{{\rm B}}+P_{3}\cos^{2}\theta_{{\rm B}}),\label{eq.I00-1_0}\\
I_{1}((00\overline{1});\omega) & = & |AB|\cos\theta_{{\rm B}}(-P_{2}\cos^{2}\theta_{{\rm B}}\pm2P_{3}\sin\theta_{{\rm B}}).\label{eq.I00-1_1}
\end{eqnarray}
Equation (\ref{eq.I00-1_0}) for $I_{0}$ is the same as Eq. (7) in
Ref.~{[}\onlinecite{Tanaka2010}{]} (see also the errata), if $|A|^{2}$
and $|B|^{2}$ are identified to be proportional to $T_{a}^{2}$ and
$T_{b}^{2}$. Equation (\ref{eq.I00-1_1}) for $I_{1}$ is similar
to Eq.~(8) in Ref.~{[}\onlinecite{Tanaka2010}{]} and to Eq.~(11)
in Refs.~{[}\onlinecite{Tanaka2012,com2}{]}, but different from
them concerning chirality dependence. In addition, the possible phase
shift $\delta$ is absent in Eq.~(6) in Ref.~{[}\onlinecite{Tanaka2010}{]}
and in Eq.~(9) in Ref.~{[}\onlinecite{Tanaka2012}{]}.

Finally, $U_{1}$, $U_{2}$, $U_{3}^{+}$ and $U_{3}^{-}$ for ${\bf G}=(00\overline{2})$
are given by the same forms as those for ${\bf G}=(00\overline{1})$
with replacing $3\Psi\mp\delta$ by $3\Psi\pm\delta$ and with reversing
the signs in the expression of $U_{2}$ for ${\bf G}=(00\overline{1})$.
Hence, the scattering intensity with $P_{1}=0$ is expressed as 
\begin{equation}
I((00\overline{2});\omega)=I_{0}((00\overline{2});\omega)+I_{1}((00\overline{2});\omega)\sin(3\Psi\pm\delta),\label{eq.I00-2}
\end{equation}
with 
\begin{eqnarray}
I_{0}((00\overline{2});\omega) & = & |B]^{2}\cos^{2}\theta_{{\rm B}}+|A|^{2}\frac{1+\sin^{2}\theta_{{\rm B}}}{2}\nonumber \\
 & \times & (1+\sin^{2}\theta_{{\rm B}}\mp2P_{2}\sin\theta_{{\rm B}}+P_{3}\cos^{2}\theta_{{\rm B}}),\label{eq.I00-2_0}\\
I_{1}((00\overline{2});\omega) & = & |AB|\cos\theta_{{\rm B}}(P_{2}\cos^{2}\theta_{{\rm B}}\pm2P_{3}\sin\theta_{{\rm B}}).\label{eq.I00-2_1}
\end{eqnarray}

\section{\label{sect.5}Bond-orbital model analysis}

We employ the bond-orbital model developed by Harrison\cite{Harrison}
to describe the electronic structures of $\alpha$-quartz and $\alpha$-berlinite.
We explain the model for $\alpha$-quartz by following Ref. {[}\onlinecite{Harrison}{]}.
The extension of the model to $\alpha$-berlinite is straightforward
by replacing one of the Si atoms by an Al atom and another one by a P atom
in the Si-O-Si bond.

\begin{figure}
\includegraphics[width=8cm]{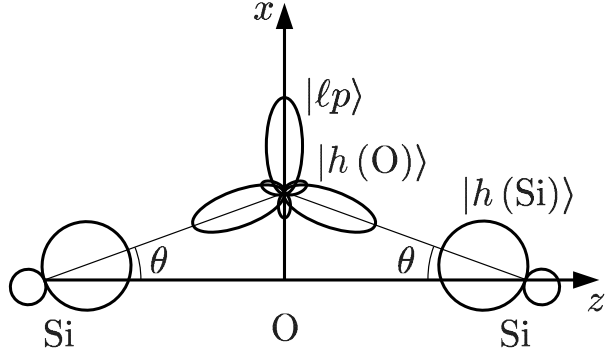}\caption{\label{fig.model} Si-O-Si bond for $\alpha$-quartz. Si atoms are
replaced by Al and P atoms for $\alpha$-berlinite.}
\end{figure}

Consider the Si-O-Si bond shown in Fig.~\ref{fig.model}. There exist
four such bonds for each Si atom. Let us introduce the Si sp$^{3}$-orbitals,
$|h({\rm Si})\rangle$. For the O atom, we construct the oxygen hybrids
to the left and right, 
\begin{eqnarray}
|h({\rm O})_{l}\rangle & = & \cos\eta|s\rangle+\sin\eta(-\cos\theta|p_{z}\rangle-\sin\theta|p_{x}\rangle),\\
|h({\rm O})_{r}\rangle & = & \cos\eta|s\rangle+\sin\eta(\cos\theta|p_{z}\rangle-\sin\theta|p_{x}\rangle),
\end{eqnarray}
where $|s\rangle$ stands for the $2s$ state, and $|p_{x}\rangle$,
$|p_{y}\rangle$, $|p_{z}\rangle$ stand for the $2p$ states of O,
respectively. Imposing the orthogonality condition $\langle h({\rm O})_{r}|h({\rm O})_{l}\rangle=0$,
we have $\cos^{2}\eta=(1-\tan^{2}\theta)/2$ with $\theta=18^{\circ}$
being the bending angle of the bond. We may call the states by sp$^{1.24}$-hybrid.
We also construct the lone-pair orbital, which we write as 
\begin{equation}
|h_{lp}\rangle=\cos\eta_{lp}|s\rangle+\sin\eta_{lp}|p_{x}\rangle.
\end{equation}
Imposing the orthogonality conditions to $|h({\rm O})_{l}\rangle$
and $|h({\rm O})_{r}\rangle$, we have $\cos^{2}\eta_{lp}=\tan^{2}\theta$.
This state as well as the oxygen $\pi$ orbital, $|\pi({\rm O})\rangle$,
are assumed to have no coupling to other states. With the bases, $|h_{1}({\rm Si})\rangle$,
$|h({\rm O})_{l}\rangle$, $|h({\rm O})_{r}\rangle$, $|h_{2}({\rm Si})\rangle$,
the Hamiltonian matrix for each bond may be represented as 
\begin{equation}
H_{{\rm bond}}=\left(\begin{array}{cccc}
\epsilon_{h1} & -V_{2} & 0 & 0\\
-V_{2} & \epsilon_{h}({\rm O)} & -V_{1-} & 0\\
0 & -V_{1-} & \epsilon_{h}({\rm O)} & -V_{2}\\
0 & 0 & -V_{2} & \epsilon_{h2}
\end{array}\right),
\end{equation}
where $\epsilon_{h1}=\epsilon_{h2}\equiv\epsilon_{h}({\rm Si})$ is
the energy of the sp$^{3}$ state of Si. The $\epsilon_{h}({\rm O})$,
the hybrid energy of O, is given by $\epsilon_{h}({\rm O})=\cos^{2}\eta\epsilon_{s}({\rm O})+\sin^{2}\eta\epsilon_{p}({\rm O})$,
with $\epsilon_{s}({\rm O})$ and $\epsilon_{p}({\rm O})$ being the
energies of $2s$ and $2p$ states of O. The covalent energy $V_{2}$
is given by 
\begin{equation}
V_{2}=-\frac{1}{2}\cos\eta V_{ss\sigma}+\left(\frac{1}{2}\sin\eta+\frac{\sqrt{3}}{2}\cos\eta\right)V_{sp\sigma}+\frac{\sqrt{3}}{2}\sin\eta V_{pp\sigma},
\end{equation}
where $V_{ss\sigma}$, $V_{sp\sigma}$, and $V_{pp\sigma}$ are the
Slater-Koster parameters.\cite{Slater1954} The $V_{1-}$ represents
the coupling between the two oxygen hybrids given by 
\begin{equation}
V_{1-}=(\epsilon_{p}({\rm O})-\epsilon_{s}({\rm O}))\cos^{2}\eta.
\end{equation}

Table \ref{table.1} shows the parameter values for $\alpha$-quartz
and $\alpha$-berlinite. Most of them are taken from Table 12-1 in
Ref. {[}\onlinecite{Harrison}{]}. The $\epsilon_{h}({\rm Al})$
and $\epsilon_{h}({\rm P})$ are taken from Table 1-1 in Ref. {[}\onlinecite{Harrison}{]}.
The covalent energy $V_{2}$ for Al-O and P-O are estimated from $V_{2}$
for Si-O by assuming the so-called $d^{-2}$ dependence.
The scattering intensities discussed later have been checked to be
insensitive to the choice of parameter values.

\begin{table*}
\caption{\label{table.1} Parameters for SiO$_{2}$ and AlPO$_{4}$. }

\begin{ruledtabular} %
\begin{tabular}{lll}
$\epsilon_{h}({\rm Si})=-9.38$\,eV  & $\epsilon_{h}({\rm O})=-24.46$\,eV  & $V_{2}({\rm Si-O})=9.47$\,eV \tabularnewline
$\epsilon_{h}({\rm Al})=-6.96$\,eV  & $\epsilon_{lp}({\rm O})=-18.55$\,eV  & $V_{2}({\rm Al-O})=8.20$\,eV \tabularnewline
$\epsilon_{h}({\rm P})=-11.96$\,eV  & $\epsilon_{\pi}({\rm O})=-16.77$\,eV  & $V_{2}({\rm P-O})=10.62$\,eV \tabularnewline
 & $\theta=18$\,deg  & $V_{1-}=7.74$\,eV \tabularnewline
\end{tabular}\end{ruledtabular} 
\end{table*}

We diagonalize the Hamiltonian matrix to obtain the eigenvalues specified
as $\epsilon_{1}^{(0)}<\epsilon_{2}^{(0)}<\epsilon_{3}^{(0)}<\epsilon_{4}^{(0)}$,
and the corresponding eigenstates denoted by $|\psi_{i}^{(0)}\rangle$
($i=1,\cdots4$). In the ground state, four electrons are occupied
on the lowest two levels, $|\psi_{1}^{(0)}\rangle$ and $|\psi_{2}^{(0)}\rangle$,
and furthermore four electrons are occupied on $|h_{lp}\rangle$ and
$|\pi({\rm O})\rangle$.

The $E1$ transition may be expressed as 
\begin{equation}
x_{\alpha}|g\rangle=\frac{1}{2}T\sum_{\sigma}\sum_{i,\ell}h_{i}^{(0)}\cos\theta_{\alpha}^{(\ell)}a_{i\sigma}^{\dagger}(\ell)a_{1s\sigma}|g\rangle,\label{eq.xg}
\end{equation}
where $x_{\alpha}$ is the dipole operator on the Si site. $T=\int R_{3p}(r)rR_{1s}(r)r^{2}{\rm d}r$
with $R_{3p}(r)$ and $R_{1s}(r)$ being the radial wave function
of $3p$ and $1s$ states. The $a_{1s\sigma}$ is the annihilation
operator of the 1s electron with spin $\sigma$, and $a_{i\sigma}^{\dagger}(\ell)$
is the creation operator of electron on orbital $|\psi_{i}^{(0)}\rangle_{\ell}$
in the $\ell$-th bond, where four bonds of Si-O-Si (or Al-O-P) are
distinguished by $\ell$. The $\cos\theta_{\alpha}^{(\ell)}$ denotes
the directional cosine of the sp$^{3}$ hybrid of the $\ell$-th bond
with respect to the $x_{\alpha}$ axis, which is determined from the
positions of oxygen atoms specified in Sec.~\ref{sect.2}. The sum
over $i$ should be restricted within the unoccupied states, \textit{i.e.}
$i=3,4$. Coefficient $h_{i}^{(0)}$ is defined by $|\psi_{i}^{(0)}\rangle_{\ell}=h_{i}^{(0)}|h_{1}\rangle_{\ell}+\cdots$,
which is independent of $\ell$.

In the intermediate state, an attractive potential from the $1s$-core
hole is working, which is known to modify the absorption coefficient
as a function of photon energy $\omega$.\cite{Taillefumier2002}
Assuming that the core-hole potential works within the Si or Al site,
we change $\langle h_{1}|H|h_{1}\rangle$ from $\epsilon_{h1}$ to
$\epsilon_{h1}+V_{{\rm core}}$. We tentatively put $V_{{\rm core}}=-6$
eV. As shown below, the dependence on polarization, chirality, and
scattering vector is unaltered by the presence of $V_{{\rm core}}$
for fixed $\omega$, although the spectral shape is modified as a
function of $\omega$. We diagonalize the Hamiltonian matrix to obtain
the eigenvalues $\epsilon_{1}^{(1)}<\epsilon_{2}^{(1)}<\epsilon_{3}^{(1)}<\epsilon_{4}^{(1)}$
and the corresponding eigenstates $|\psi_{i}^{(1)}\rangle$ ($i=1,\cdots4$).
The intermediate state is constructed by distributing five electrons
on these energy levels in one of the four bonds (other three bonds
are occupied four electrons).

The dominant contributions come from the intermediate states that
four electrons occupy lowest two states and one electron occupies
higher state. Thus, neglecting the so-called shake-up states in the
intermediate state, we consider the overlap between the intermediate
state and the ground state, and obtain the local dipole-dipole correlation
function, 
\begin{equation}
[\hat{\rho}_{0}^{(\pm)}(\epsilon)]_{\alpha,\beta}=\frac{1}{2}T^{2}Q(\epsilon)\sum_{\ell}\cos\theta_{\alpha}^{(\ell)}\cos\theta_{\beta}^{(\ell)},\label{eq.rho1}
\end{equation}
where 
\begin{align}
Q(\epsilon) & =\left|O_{g}\right|^{14}\Bigl\{|O_{3,3}h_{3}^{(0)}+O_{3,4}h_{4}^{(0)}|^{2}\delta(\epsilon-\epsilon_{{\rm core}}-\epsilon_{{\rm ex}}^{0}-\epsilon_{3}^{(1)})\nonumber \\
 & \phantom{=\left|O_{g}\right|^{14}\Bigl\{}+|O_{4,3}h_{3}^{(0)}+O_{4,4}h_{4}^{(0)}|^{2}\delta(\epsilon-\epsilon_{{\rm core}}-\epsilon_{{\rm ex}}^{0}-\epsilon_{4}^{(1)})\Bigr\},
\end{align}
with 
\begin{eqnarray}
O_{g} & = & \left|\begin{array}{cc}
\langle\psi_{1}^{(1)}|\psi_{1}^{(0)}\rangle & \langle\psi_{1}^{(1)}|\psi_{2}^{(0)}\rangle\\
\langle\psi_{2}^{(1)}|\psi_{1}^{(0)}\rangle & \langle\psi_{2}^{(1)}|\psi_{2}^{(0)}\rangle
\end{array}\right|,\\
O_{n,n'} & = & \left|\begin{array}{ccc}
\langle\psi_{1}^{(1)}|\psi_{1}^{(0)}\rangle & \langle\psi_{1}^{(1)}|\psi_{2}^{(0)}\rangle & \langle\psi_{1}^{(1)}|\psi_{n'}^{(0)}\rangle\\
\langle\psi_{2}^{(1)}|\psi_{1}^{(0)}\rangle & \langle\psi_{2}^{(1)}|\psi_{2}^{(0)}\rangle & \langle\psi_{2}^{(1)}|\psi_{n'}^{(0)}\rangle\\
\langle\psi_{n}^{(1)}|\psi_{1}^{(0)}\rangle & \langle\psi_{n}^{(1)}|\psi_{2}^{(0)}\rangle & \langle\psi_{n}^{(1)}|\psi_{n'}^{(0)}\rangle
\end{array}\right|,\\
\epsilon_{{\rm ex}}^{0} & = & 8(\epsilon_{1}^{(1)}+\epsilon_{2}^{(1)}-\epsilon_{1}^{(0)}-\epsilon_{2}^{(0)}).
\end{eqnarray}
We notice from Eq.~(\ref{eq.rho1}) that all the components $a(\epsilon)$,
$b(\epsilon)$, $c(\epsilon)$, $d(\epsilon)$, are proportional to
$Q(\epsilon)$, and that their relative ratios are determined by the
directions of Si-O or Al-O bonds. Hence we obtain $A(\omega)$ and
$B(\omega)$ from Eqs.~(\ref{eq.A}) and (\ref{eq.B}), which are
proportional to 
\[
T^{2}\int\frac{Q(\epsilon)}{\omega-\epsilon+i\Gamma}{\rm d}\epsilon.
\]
Accordingly $B^{*}(\omega)A(\omega)$ is real for both $\alpha$-quartz
and $\alpha$-berlinite, indicating that the phase shift $\delta$
is generally $0$ or $\pi$, independent of $\omega$. These simple
result would be modified, if the coupling between bonds is taken into
account.

The absorption coefficient $C(\omega)$ for photon energy $\omega$
may be expressed as 
\begin{equation}
C(\omega)\propto\sum_{\alpha}\sum_{n}|\langle n|x_{\alpha}|g\rangle|^{2}\frac{\Gamma/\pi}{(\omega+\epsilon_{g}-\epsilon_{n})^{2}+\Gamma^{2}}.
\end{equation}
Therefore we obtain from Eqs.~(\ref{eq.local.density}) and (\ref{eq.rho1})
\begin{equation}
C(\omega)\propto\int{\rm Tr}\,\hat{\rho}(j;\epsilon)\frac{\Gamma/\pi}{(\omega-\epsilon)^{2}+\Gamma^{2}}{\rm d}\epsilon=2T^{2}\frac{\Gamma}{\pi}\int\frac{Q(\epsilon)}{(\omega-\epsilon)^{2}+\Gamma^{2}}{\rm d}\epsilon.\label{eq.absA}
\end{equation}

\section{\label{sect.6}Calculated results in comparison with experiments}

\subsection{$\alpha$-quartz}

\begin{figure}
\includegraphics[width=8cm]{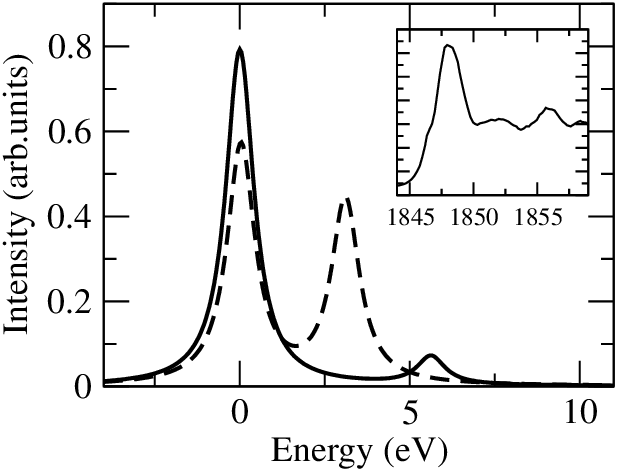}\caption{\label{fig.quartz.abs} Absorption coefficient as a function of photon
energy for Si-$K$-edge in $\alpha$-quartz. The solid and broken
lines are the calculated results with and without the core-hole potential.
The origin of photon energy is set to the main peak position. Inset
displays the experimental absorption data taken from Ref.{[}\onlinecite{Tanaka2008}{]}. }
\end{figure}

Figure \ref{fig.quartz.abs} shows the absorption coefficient calculated
from Eq.~(\ref{eq.absA}), in comparison with the experiment.\cite{Tanaka2008}
The first and second peaks arise from the transition from the $1s$
state to the unoccupied state $|\psi_{3}^{(0)}\rangle$, and to 
$|\psi_{4}^{(0)}\rangle$, respectively. The core-hole potential makes 
the first peak intensity increase.
In actuality, there exist more $p$-symmetric states forming band states,
which make the intensities spread above the $K$ edge.
The core-hole potential gives rise to the intensity transfer from
the high energy region, resulting in sharpening the peak at the edge,
as shown in Ref.~{[}\onlinecite{Taillefumier2002}{]}. 
The states constituting the peak are close to the localized states constructed
by the sp$^3$-orbitals. Therefore the present model could describe rather well
the states of the peak at the edge. 
Note that there exists a shoulder in the experimental curve of the absorption 
coefficient.\cite{Tanaka2008} Such a peak is, however, not seen in the 
experimental and the theoretical curves 
in Ref.~{[}\onlinecite{Taillefumier2002}{]}. 
The origin of this shoulder is not clear.

Note that the RXS intensity
as a function of $\omega$ is proportional to 
\[
\left\{ \int\frac{Q(\epsilon)}{\omega-\epsilon+i\Gamma}{\rm d}\epsilon\right\} ^{*}\left\{ \int\frac{Q(\epsilon)}{\omega-\epsilon+i\Gamma}{\rm d}\epsilon\right\} .
\]
Although it is not shown here, this quantity has a large peak at $\omega$
giving the large peak in the absorption coefficient, in agreement
with the experiment.\cite{Tanaka2008}

\begin{figure}
\includegraphics[width=8cm]{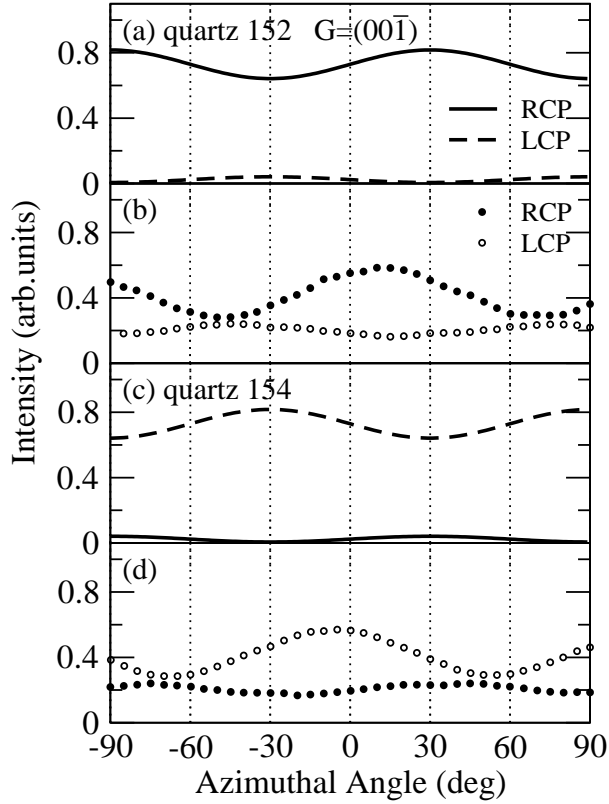}\caption{\label{fig.azim.quartz} RXS intensity from $\alpha$-quartz as a
function of azimuthal angle $\Psi$ for ${\bf G}=(00\overline{1})$.
The $\omega$ is fixed at the value giving the absorption peak. $P_{1}=0$,
$P_{3}=-0.31$, $P_{2}=0.95$ (RCP), and $P_{2}=-0.95$ (LCP). Panels
(b) and (d) show the experimental data in No.152 and No.154, respectively,
which are taken from Ref.~{[}\onlinecite{Tanaka2008}{]}. }
\end{figure}

We concentrate our attention on the spectra at the $\omega$ giving
the main absorption peak in the following. Figure \ref{fig.azim.quartz}
shows the RXS intensity as a function of azimuthal angle $\Psi$ for
${\bf G}=(00\overline{1})$. Although ${\bf G}$ is defined by ${\bf k}_{i}-{\bf k}_{f}$
in the experiment, which is opposite to ours (see the errata in Ref.~{[}\onlinecite{Tanaka2008}{]}),
we present ${\bf G}$'s in our definition. The Stokes parameters are
set to be $P_{1}=0$, $P_{2}=\pm0.95$, $P_{3}=-0.31$ in accordance
with the experiment.\cite{Tanaka2008} For ${\bf G}=(00\overline{1})$,
$\sin\theta_{{\rm B}}=0.625$, $\cos\theta_{{\rm B}}=0.781$. The
calculated intensities are larger for RCP ($P_{2}=0.95$) than for
LCP ($P_{2}=-0.95$) in No.152, while the former is smaller than the
latter in No.154, consistent with the experimental curves shown
in panels (b) and (d).\cite{Tanaka2008} The intensities for LCP in
No.152 as well as for RCP in No.154 are, however, too small in comparison
with the experiment. We hope that the absorption correction to the experimental
data, if it were not made yet, as well as a careful subtraction of the 
background might improve the discrepancy. As regards
the oscillation terms, we notice from the general expressions (Eqs.~(\ref{eq.I00-1})-(\ref{eq.I00-1_1}))
that they take the form of $-a_{1}\sin(3\Psi-\delta)$ for RCP in
No.152, and $a_{1}\sin(3\Psi+\delta)$ for LCP in No.154 with $a_{1}$
a positive number, and that they take the form of $b_{1}\sin(3\Psi-\delta)$
for LCP in No.152, and $-b_{1}\sin(3\Psi+\delta)$ for RCP in No.154
with $b_{1}$ a positive number. The amplitude $a_{1}$ is much larger
than the amplitude $b_{1}$; their ratio is given by $b_{1}/a_{1}=(|P_{2}|\cos^{2}\theta_{{\rm B}}+2P_{3}\sin\theta_{{\rm B}})/(|P_{2}|\cos^{2}\theta_{{\rm B}}-2P_{3}\sin\theta_{{\rm B}})\sim0.2$,
which is independent of the model. The experimental curves seem to
belong to these forms with the phase shift $\delta=2\pi/3\sim\pi$.
The present calculation within the bond-orbital model gives the phase
shift $\delta=\pi$.

\begin{figure}
\includegraphics[width=8cm]{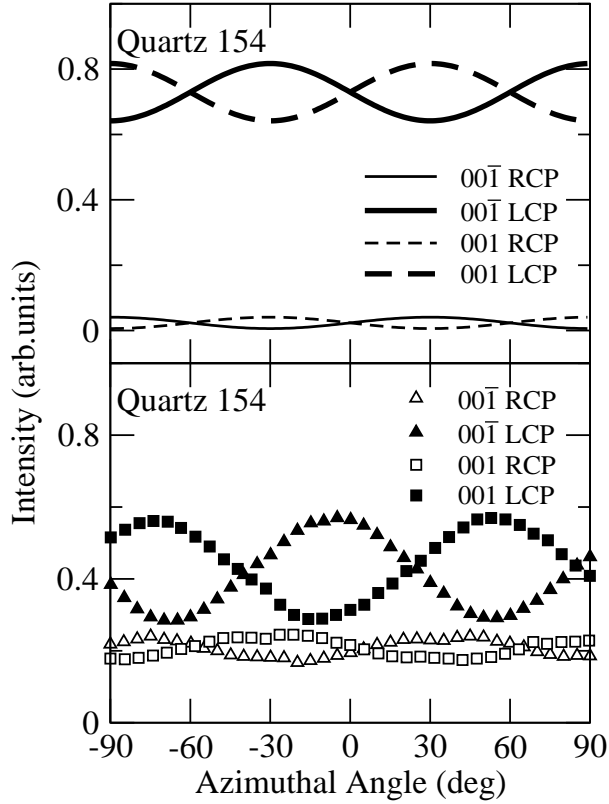}\caption{\label{fig.azim.quartz.001-1} RXS intensity from $\alpha$-quartz
of No.154, for ${\bf G}=(001)$ and $(00\overline{1})$. The $\omega$
is fixed at the value giving the absorption peak. $P_{1}=0$, $P_{3}=-0.31$,
$P_{2}=0.95$ (RCP), and $P_{2}=-0.95$ (LCP). Lower panel shows the
experimental curves reproduced from Ref.~{[}\onlinecite{Tanaka2008}{]}. }
\end{figure}

Figure \ref{fig.azim.quartz.001-1} shows the RXS intensities for
both ${\bf G}=(001)$ and ${\bf G}=(00\overline{1})$ in No. 154.
According to the general expressions (Eqs.~(\ref{eq.I001}) and (\ref{eq.I00-1})),
the oscillation terms are proportional to $\sin(3\Psi+\delta)$ for
both ${\bf G}=(001)$ and $(00\overline{1})$, regardless of RCP or
LCP, in No.154. This means that all curves have to be maximum or minimum
at the same $\Psi$-values. This requirement seems not to be satisfied
in the experimental curves, where the maximum and minimum positions
are somewhat different, indicating that the phase does not seems to
has a definite value. 

\subsection{$\alpha$-berlinite}

\begin{figure}
\includegraphics[width=8cm]{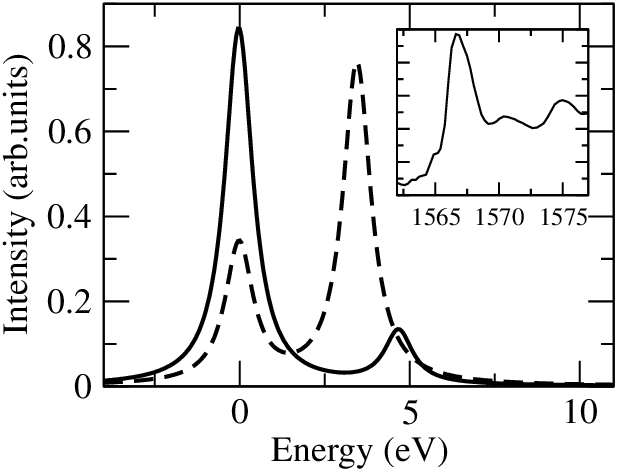}\caption{\label{fig.berlinite.abs} Absorption coefficient as a function of
photon energy for Al-$K$-edge in $\alpha$-berlinite. The solid and
broken lines are the calculated results with and without the core-hole
potential. The origin of photon energy is set to the main peak position.
The inset displays the experimental absorption data taken from Ref.~{[}\onlinecite{Tanaka2010}{]}. }
\end{figure}

Figure \ref{fig.berlinite.abs} shows the absorption coefficient calculated
from Eq.~(\ref{eq.absA}), in comparison with the experiment. \cite{Tanaka2010}
Without taking account of the core-hole potential, we have the second-peak
intensity larger than the first-peak one, which in fact is different
from $\alpha$-quartz. The core-hole potential makes the first-peak
intensity larger than the second one, in agreement with the experiment.\cite{Tanaka2010}

\begin{figure}
\includegraphics[width=7cm]{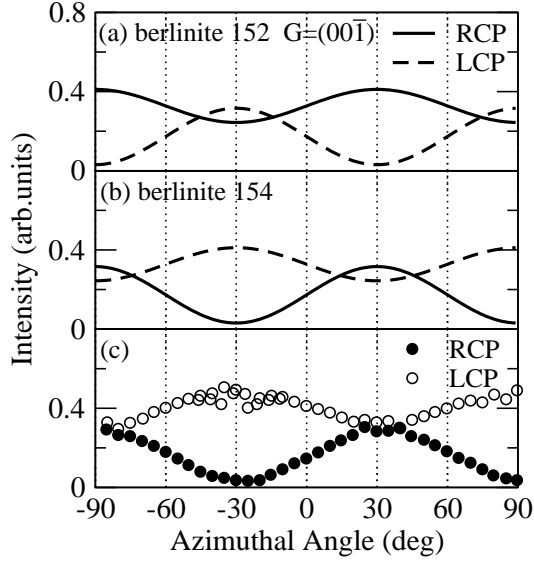}\caption{\label{fig.azim.berlinite_00-1} RXS intensity from $\alpha$-berlinite
as a function of azimuthal angle $\Psi$, for ${\bf G}=(00\overline{1})$.
The $\omega$ is fixed at the value giving the absorption peak. $P_{1}=0$,
$P_{3}=0.30$, $P_{2}=0.95$ (RCP), and $P_{2}=-0.95$ (LCP). Panel
(c) shows the experimental data for No.154, taken from Ref.~{[}\onlinecite{Tanaka2010}{]}. }
\end{figure}

We concentrate our attention on the spectra at the $\omega$ giving
the absorption peak. Figure \ref{fig.azim.berlinite_00-1} shows the
RXS intensity for ${\bf G}=(00\overline{1})$, where $\sin\theta_{{\rm B}}=0.361$,
$\cos\theta_{{\rm B}}=0.932$. The Stokes parameters are set to be
$P_{1}=0$, $P_{2}=\pm0.95$, $P_{3}=0.30$, in accordance with the
experiment,\cite{Tanaka2010} where the value of $P_{3}$ here is
opposite in sign to the case of $\alpha$-quartz.\cite{Tanaka2010}
The average intensity for RCP is larger than for LCP in No.152, while
the former is smaller than the latter in No.154, which dependence
is the same as in $\alpha$-quartz. The oscillation terms take the
same forms as for $\alpha$-quartz; the ratio of the amplitudes are
given by $b_{1}/a_{1}\sim1.71$. The present calculation gives the
phase shift $\delta=\pi$. The calculated curves are in good agreement
with the experimental curves shown in panel (c).\cite{Tanaka2010}
We would like to emphasize that there exist no adjustable parameter
in the present calculation.

\begin{figure}
\includegraphics[width=7cm]{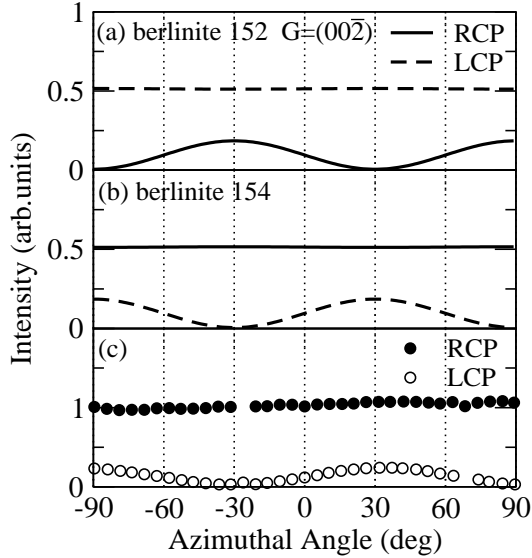}\caption{\label{fig.azim.berlinite_00-2} RXS intensity from $\alpha$-berlinite
as a function of $\Psi$, for ${\bf G}=(00\overline{2})$. The $\omega$
is fixed at the value giving the absorption peak. $P_{1}=0$, $P_{3}=0.30$,
$P_{2}=0.95$ (RCP), and $P_{2}=-0.95$ (LCP). Panel (c) shows the
experimental data for No.154, taken from Ref.~{[}\onlinecite{Tanaka2010}{]}. }
\end{figure}

Figure \ref{fig.azim.berlinite_00-2} shows the RXS intensity for
${\bf G}=(00\overline{2})$, where $\sin\theta_{{\rm B}}=0.723$,
$\cos\theta_{{\rm B}}=0.691$. The Stokes parameters are the same
as for ${\bf G}=(00\overline{1})$. The intensity for LCP is larger
than that for RCP in No.152, while the former is smaller than the
latter in No.154. This dependence on polarization is opposite to that
in ${\bf G}=(00\overline{1})$. According to the general expressions
(Eqs.~(\ref{eq.I00-2}) and (\ref{eq.I00-2_1})), the oscillation
terms take the form of $a_{2}\sin(3\Psi+\delta)$ for RCP in No.152,
and $-a_{2}\sin(3\Psi-\delta)$ for LCP in No.154 with $a_{2}$ a
positive number, while they take the form of $-b_{2}\sin(3\Psi+\delta)$
for LCP in No.152, and $b_{2}\sin(3\Psi-\delta)$ for RCP in No.154
with $b_{2}$ a positive number. The ratio of the amplitude is given
by $b_{2}/a_{2}=(|P_{2}|\cos^{2}\theta_{{\rm B}}-2P_{3}\sin\theta_{{\rm B}})/(|P_{2}|\cos^{2}\theta_{{\rm B}}+2P_{3}\sin\theta_{{\rm B}})\sim0.02$,
corresponding to the nearly flat curves for LCP in No.152 and for
RCP in No.154. The calculated curves are in good agreement with the
experimental curves shown in Panel (c).

\section{\label{sect.7} Concluding Remarks}

We have analyzed the RXS spectra at Si and Al K-edges on forbidden
spots in chiral materials, $\alpha$-quartz and $\alpha$-berlinite.
Summing up the local scattering matrices which satisfy the symmetry
requirement, we have derived the general expression of scattering
matrix on forbidden spots with E1-E1 process. We have obtained the
oscillation term as a function of azimuthal angle, which has the phase
shift and the amplitude with chirality dependence different from the
previous studies. We have evaluated the parameters undetermined by
the symmetry argument alone on the basis of the underlying electronic
structures given by the bond-orbital model. With such evaluation,
the scattering matrix is completely determined. We have calculated
the spectra depending on polarization, chirality, and scattering vector
in agreement with the experiments, although some discrepancies remain
in the average intensities and the phase shift of oscillation in $\alpha$-quartz.
The spectra for $\alpha$-berlinite have reproduced particularly well
the experiment. It should be emphasized that this result is obtained
without adjustable parameters.

We have obtained the phase shift of oscillation $\pi$ in both $\alpha$-quartz
and $\alpha$-berlinite. For $\alpha$-berlinite, the phase shift
$\pi$ is consistent with the experiment. Note that the same phase
shift is also observed in the chiral metal Te.\cite{Tanaka2010-2,com1}
On the other hand, the situation of $\alpha$-quartz is different,
where the phase shift looks deviating from $\pi$ with depending on
polarization and scattering vector (see the lower panel in Fig.~\ref{fig.azim.quartz.001-1}).
The present formula based on the symmetry requirement does not allow
such dependence but allows dependence on photon energy. Since both
materials belong to the same covalent-bonding with similar electronic
structures, such difference in the spectra between $\alpha$-quartz
and $\alpha$-berlinite is puzzling to us. To clarify the origin of
the phase shift and to obtain better agreement with experiment, it
may be necessary to take account of the coupling between the bond
orbitals, or more precisely, the band effects. 
It might be necessary to consider the absorption correction to the 
experimental curves, if it were not made, for more quantitative comparison.

Finally we comment on the effect of the E1-E2 process. Since the space
inversion symmetry is broken around Si sites in $\alpha$-quartz and
around Al sites in $\alpha$-berlinite, the $p$-symmetric states
could mix with the $d$-symmetric states, and therefore the second-order
process using both the E1 transition and the electric-quadrupole (E2)
transition could take place. This E1-E2 process is known to be important
for x-ray absorption as well as RXS at the pre-K-edge in the transition-metal
compounds such as $\alpha$-Fe$_{2}$O$_{3}$, \cite{Finkelstein1992}
K$_{2}$CrO$_{4}$,\cite{Templeton1994} 
magnetite,\cite{Matsubara2005,Igarashi2008} and GaFeO$_{3}$.\cite{Kubota2004,Arima2005,Matteo2006,Igarashi2010}
These spectra have been analyzed based on the symmetry or by taking
account of the microscopic electronic structures.\cite{Matteo2006,Igarashi2008}
Now, for $\alpha$-quartz, it has been proposed\cite{Tanaka2008,Lovesey2008,Tanaka2010,Tanaka2012}
that the phase shift is brought about by adding the scattering amplitude
coming from the E1-E2 process to that from the E1-E1 process. Since
the former contribution is expected to be more than one order of magnitude
smaller than the the latter, we think it unlikely to expect the substantial
phase shift deviation from this mechanism.
\begin{acknowledgments}
This work was partially supported by a Grant-in-Aid for Scientific
Research from the Ministry of Education, Culture, Sports, Science
and Technology of the Japanese Government.
\end{acknowledgments}

 \bibliographystyle{apsrev}
\bibliography{paper}

\begin{thebibliography}{27}
\expandafter\ifx\csname natexlab\endcsname\relax\def\natexlab#1{#1}\fi
\expandafter\ifx\csname bibnamefont\endcsname\relax
  \def\bibnamefont#1{#1}\fi
\expandafter\ifx\csname bibfnamefont\endcsname\relax
  \def\bibfnamefont#1{#1}\fi
\expandafter\ifx\csname citenamefont\endcsname\relax
  \def\citenamefont#1{#1}\fi
\expandafter\ifx\csname url\endcsname\relax
  \def\url#1{\texttt{#1}}\fi
\expandafter\ifx\csname urlprefix\endcsname\relax\def\urlprefix{URL }\fi
\providecommand{\bibinfo}[2]{#2}
\providecommand{\eprint}[2][]{\url{#2}}

\bibitem[{Mas()}]{Mason1982}
\bibinfo{note}{For a historical account, see for example, S. F. Mason,
  \textit{Molecular Optical Activity and the Chiral Discriminations} (Cambridge
  University Press, 1982).}

\bibitem[{\citenamefont{Bijvoet et~al.}(1951)\citenamefont{Bijvoet, Peerdeman,
  and van Bommel}}]{Bijvoet1951}
\bibinfo{author}{\bibfnamefont{J.~M.} \bibnamefont{Bijvoet}},
  \bibinfo{author}{\bibfnamefont{A.~F.} \bibnamefont{Peerdeman}},
  \bibnamefont{and} \bibinfo{author}{\bibfnamefont{J.~A.} \bibnamefont{van
  Bommel}}, \bibinfo{journal}{Nature} \textbf{\bibinfo{volume}{168}},
  \bibinfo{pages}{271} (\bibinfo{year}{1951}).

\bibitem[{\citenamefont{de~Vries}(1958)}]{Vries1958}
\bibinfo{author}{\bibfnamefont{A.}~\bibnamefont{de~Vries}},
  \bibinfo{journal}{Nature} \textbf{\bibinfo{volume}{181}},
  \bibinfo{pages}{1193} (\bibinfo{year}{1958}).

\bibitem[{\citenamefont{Tanaka et~al.}(2012{\natexlab{a}})\citenamefont{Tanaka,
  Takeuchi, Lovesey, Knight, Chainani, Tanaka, Oura, Senba, Ohashi, and
  Shin}}]{Tanaka2008}
\bibinfo{author}{\bibfnamefont{Y.}~\bibnamefont{Tanaka}},
  \bibinfo{author}{\bibfnamefont{T.}~\bibnamefont{Takeuchi}},
  \bibinfo{author}{\bibfnamefont{S.~W.} \bibnamefont{Lovesey}},
  \bibinfo{author}{\bibfnamefont{K.~S.} \bibnamefont{Knight}},
  \bibinfo{author}{\bibfnamefont{A.}~\bibnamefont{Chainani}},
  \bibinfo{author}{\bibfnamefont{Y.}~\bibnamefont{Tanaka}},
  \bibinfo{author}{\bibfnamefont{M.}~\bibnamefont{Oura}},
  \bibinfo{author}{\bibfnamefont{Y.}~\bibnamefont{Senba}},
  \bibinfo{author}{\bibfnamefont{H.}~\bibnamefont{Ohashi}}, \bibnamefont{and}
  \bibinfo{author}{\bibfnamefont{S.}~\bibnamefont{Shin}},
  \bibinfo{journal}{Phys.\ Rev.\ Lett.} \textbf{\bibinfo{volume}{100}},
  \bibinfo{pages}{145502(2008); \textit{ibid.} \textbf{108}, 019901(E)}
  (\bibinfo{year}{2012}{\natexlab{a}}).

\bibitem[{\citenamefont{Tanaka et~al.}(2011)\citenamefont{Tanaka, Kojima,
  Takata, Chainami, Lovesey, Knight, Takeuchi, Oura, Senba, Ohashi
  et~al.}}]{Tanaka2010}
\bibinfo{author}{\bibfnamefont{Y.}~\bibnamefont{Tanaka}},
  \bibinfo{author}{\bibfnamefont{T.}~\bibnamefont{Kojima}},
  \bibinfo{author}{\bibfnamefont{Y.}~\bibnamefont{Takata}},
  \bibinfo{author}{\bibfnamefont{A.}~\bibnamefont{Chainami}},
  \bibinfo{author}{\bibfnamefont{S.~W.} \bibnamefont{Lovesey}},
  \bibinfo{author}{\bibfnamefont{K.~S.} \bibnamefont{Knight}},
  \bibinfo{author}{\bibfnamefont{T.}~\bibnamefont{Takeuchi}},
  \bibinfo{author}{\bibfnamefont{M.}~\bibnamefont{Oura}},
  \bibinfo{author}{\bibfnamefont{Y.}~\bibnamefont{Senba}},
  \bibinfo{author}{\bibfnamefont{H.}~\bibnamefont{Ohashi}},
  \bibnamefont{et~al.}, \bibinfo{journal}{Phys.\ Rev.\ B}
  \textbf{\bibinfo{volume}{81}}, \bibinfo{pages}{144104(2010); \textit{ibid.}
  \textbf{84}, 219905(E)} (\bibinfo{year}{2011}).

\bibitem[{\citenamefont{Templeton and Templeton}(1982)}]{Templeton1982}
\bibinfo{author}{\bibfnamefont{D.~H.} \bibnamefont{Templeton}}
  \bibnamefont{and} \bibinfo{author}{\bibfnamefont{L.~K.}
  \bibnamefont{Templeton}}, \bibinfo{journal}{Acta Crystallogr.\ A}
  \textbf{\bibinfo{volume}{38}}, \bibinfo{pages}{62} (\bibinfo{year}{1982}).

\bibitem[{\citenamefont{Dmitrienko}(1983)}]{Dmitrienko1983}
\bibinfo{author}{\bibfnamefont{V.~E.} \bibnamefont{Dmitrienko}},
  \bibinfo{journal}{Acta Crystallogr.\ A} \textbf{\bibinfo{volume}{39}},
  \bibinfo{pages}{29} (\bibinfo{year}{1983}).

\bibitem[{\citenamefont{Dmitrienko et~al.}(2005)\citenamefont{Dmitrienko,
  Ishida, Kirfel, and Ovchinnikova}}]{Dmitrienko2005}
\bibinfo{author}{\bibfnamefont{V.~E.} \bibnamefont{Dmitrienko}},
  \bibinfo{author}{\bibfnamefont{K.}~\bibnamefont{Ishida}},
  \bibinfo{author}{\bibfnamefont{A.}~\bibnamefont{Kirfel}}, \bibnamefont{and}
  \bibinfo{author}{\bibfnamefont{E.~N.} \bibnamefont{Ovchinnikova}},
  \bibinfo{journal}{Acta Crystallogr.\ A} \textbf{\bibinfo{volume}{61}},
  \bibinfo{pages}{481} (\bibinfo{year}{2005}).

\bibitem[{\citenamefont{Lovesey et~al.}(2008)\citenamefont{Lovesey, Balcar, and
  Tanaka}}]{Lovesey2008}
\bibinfo{author}{\bibfnamefont{S.~W.} \bibnamefont{Lovesey}},
  \bibinfo{author}{\bibfnamefont{E.}~\bibnamefont{Balcar}}, \bibnamefont{and}
  \bibinfo{author}{\bibfnamefont{Y.}~\bibnamefont{Tanaka}},
  \bibinfo{journal}{J.\ Phys.: Condens.\ Matter} \textbf{\bibinfo{volume}{20}},
  \bibinfo{pages}{272201} (\bibinfo{year}{2008}).

\bibitem[{\citenamefont{Tanaka et~al.}(2012{\natexlab{b}})\citenamefont{Tanaka,
  Collins, Lovesey, Matsumami, Moriwaki, and Shin}}]{Tanaka2010-2}
\bibinfo{author}{\bibfnamefont{Y.}~\bibnamefont{Tanaka}},
  \bibinfo{author}{\bibfnamefont{S.~P.} \bibnamefont{Collins}},
  \bibinfo{author}{\bibfnamefont{S.~W.} \bibnamefont{Lovesey}},
  \bibinfo{author}{\bibfnamefont{M.}~\bibnamefont{Matsumami}},
  \bibinfo{author}{\bibfnamefont{M.}~\bibnamefont{Moriwaki}}, \bibnamefont{and}
  \bibinfo{author}{\bibfnamefont{S.}~\bibnamefont{Shin}}, \bibinfo{journal}{J.\
  Phys.: Condens.\ Matter} \textbf{\bibinfo{volume}{22}},
  \bibinfo{pages}{1220(2010); \textit{ibid.} \textbf{24}, 159905(E)}
  (\bibinfo{year}{2012}{\natexlab{b}}).

\bibitem[{\citenamefont{Tanaka and Lovesey}(2012)}]{Tanaka2012}
\bibinfo{author}{\bibfnamefont{Y.}~\bibnamefont{Tanaka}} \bibnamefont{and}
  \bibinfo{author}{\bibfnamefont{S.~W.} \bibnamefont{Lovesey}},
  \bibinfo{journal}{Eur.\ Phys.\ J.\ Special Topics}
  \textbf{\bibinfo{volume}{208}}, \bibinfo{pages}{69} (\bibinfo{year}{2012}).

\bibitem[{Har()}]{Harrison}
\bibinfo{note}{W. A. Harison, \textit{Elementary Electronic Structure (Revised
  Edition)} (World Scientific, 2004). References are therein.}

\bibitem[{\citenamefont{Page and Donnay}(1976)}]{Page1976}
\bibinfo{author}{\bibfnamefont{Y.~L.} \bibnamefont{Page}} \bibnamefont{and}
  \bibinfo{author}{\bibfnamefont{G.}~\bibnamefont{Donnay}},
  \bibinfo{journal}{Acta.\ Cryst.\ B} \textbf{\bibinfo{volume}{32}},
  \bibinfo{pages}{2456} (\bibinfo{year}{1976}).

\bibitem[{\citenamefont{Muraoka and Kihara}(1997)}]{Muraoka1997}
\bibinfo{author}{\bibfnamefont{Y.}~\bibnamefont{Muraoka}} \bibnamefont{and}
  \bibinfo{author}{\bibfnamefont{K.}~\bibnamefont{Kihara}},
  \bibinfo{journal}{Phys.\ Chem.\ Materials} \textbf{\bibinfo{volume}{24}},
  \bibinfo{pages}{243} (\bibinfo{year}{1997}).

\bibitem[{Lan()}]{Landau}
\bibinfo{note}{V. B berestetskii, E. M. Lifshitz, and L. P. Pitaevskii,
  \textit{Quantum Electrodynamics} (Butterworth-Heinemann, 1982), Sec. 8.}

\bibitem[{com({\natexlab{a}})}]{com2}
\bibinfo{note}{$T_{\alpha}$ and $T_{\beta}$ in Ref.~[\onlinecite{Tanaka2012}]
  are $T_{\alpha}$ and $T_{b}\cos\theta_{\rm B}$ in
  Ref.~[\onlinecite{Tanaka2010}], respectively.}

\bibitem[{\citenamefont{Slater and Koster}(1954)}]{Slater1954}
\bibinfo{author}{\bibfnamefont{J.~C.} \bibnamefont{Slater}} \bibnamefont{and}
  \bibinfo{author}{\bibfnamefont{G.~F.} \bibnamefont{Koster}},
  \bibinfo{journal}{Phys.\ Rev.} \textbf{\bibinfo{volume}{94}},
  \bibinfo{pages}{1498} (\bibinfo{year}{1954}).

\bibitem[{\citenamefont{Taillefumier et~al.}(2002)\citenamefont{Taillefumier,
  Cabareti, Flank, and Mauri}}]{Taillefumier2002}
\bibinfo{author}{\bibfnamefont{M.}~\bibnamefont{Taillefumier}},
  \bibinfo{author}{\bibfnamefont{D.}~\bibnamefont{Cabareti}},
  \bibinfo{author}{\bibfnamefont{M.}~\bibnamefont{Flank}}, \bibnamefont{and}
  \bibinfo{author}{\bibfnamefont{F.}~\bibnamefont{Mauri}},
  \bibinfo{journal}{Phys.\ Rev. B} \textbf{\bibinfo{volume}{66}},
  \bibinfo{pages}{195107} (\bibinfo{year}{2002}).

\bibitem[{com({\natexlab{b}})}]{com1}
\bibinfo{note}{In Ref.~[\onlinecite{Tanaka2010-2}], the origin of $\Psi$ is
  defined such that the scattering plane includes $a^{*}$, which is
  $30^{\circ}$ different from the definition in the experiments of
  $\alpha$-quartz and $\alpha$-berlinite.}

\bibitem[{\citenamefont{Finkelstein et~al.}(1992)\citenamefont{Finkelstein, O,
  and Shastri}}]{Finkelstein1992}
\bibinfo{author}{\bibfnamefont{K.~D.} \bibnamefont{Finkelstein}},
  \bibinfo{author}{\bibfnamefont{S.}~\bibnamefont{O}}, \bibnamefont{and}
  \bibinfo{author}{\bibfnamefont{S.}~\bibnamefont{Shastri}},
  \bibinfo{journal}{Phys.\ Rev.\ Lett.} \textbf{\bibinfo{volume}{69}},
  \bibinfo{pages}{1612} (\bibinfo{year}{1992}).

\bibitem[{\citenamefont{Templeton and Templeton}(1994)}]{Templeton1994}
\bibinfo{author}{\bibfnamefont{D.~H.} \bibnamefont{Templeton}}
  \bibnamefont{and} \bibinfo{author}{\bibfnamefont{L.~K.}
  \bibnamefont{Templeton}}, \bibinfo{journal}{Phys.\ Rev.\ B}
  \textbf{\bibinfo{volume}{49}}, \bibinfo{pages}{14850} (\bibinfo{year}{1994}).

\bibitem[{\citenamefont{Matsubara et~al.}(2005)\citenamefont{Matsubara,
  Shimada, Arima, Taguchi, and Tokura}}]{Matsubara2005}
\bibinfo{author}{\bibfnamefont{M.}~\bibnamefont{Matsubara}},
  \bibinfo{author}{\bibfnamefont{Y.}~\bibnamefont{Shimada}},
  \bibinfo{author}{\bibfnamefont{T.}~\bibnamefont{Arima}},
  \bibinfo{author}{\bibfnamefont{Y.}~\bibnamefont{Taguchi}}, \bibnamefont{and}
  \bibinfo{author}{\bibfnamefont{Y.}~\bibnamefont{Tokura}},
  \bibinfo{journal}{Phys.\ Rev.\ B} \textbf{\bibinfo{volume}{72}},
  \bibinfo{pages}{220404} (\bibinfo{year}{2005}).

\bibitem[{\citenamefont{Igarashi and Nagao}(2008)}]{Igarashi2008}
\bibinfo{author}{\bibfnamefont{J.}~\bibnamefont{Igarashi}} \bibnamefont{and}
  \bibinfo{author}{\bibfnamefont{T.}~\bibnamefont{Nagao}},
  \bibinfo{journal}{J.\ Phys.\ Soc.\ Jpn.} \textbf{\bibinfo{volume}{77}},
  \bibinfo{pages}{084706} (\bibinfo{year}{2008}).

\bibitem[{\citenamefont{Kubota et~al.}(2004)\citenamefont{Kubota, Arima,
  Kaneko, He, Yu, and Tokura}}]{Kubota2004}
\bibinfo{author}{\bibfnamefont{M.}~\bibnamefont{Kubota}},
  \bibinfo{author}{\bibfnamefont{T.}~\bibnamefont{Arima}},
  \bibinfo{author}{\bibfnamefont{Y.}~\bibnamefont{Kaneko}},
  \bibinfo{author}{\bibfnamefont{J.~P.} \bibnamefont{He}},
  \bibinfo{author}{\bibfnamefont{X.~Z.} \bibnamefont{Yu}}, \bibnamefont{and}
  \bibinfo{author}{\bibfnamefont{Y.}~\bibnamefont{Tokura}},
  \bibinfo{journal}{Phys.\ Rev.\ Lett.} \textbf{\bibinfo{volume}{92}},
  \bibinfo{pages}{137401} (\bibinfo{year}{2004}).

\bibitem[{\citenamefont{Arima et~al.}(2005)\citenamefont{Arima, Jung,
  Matsubara, Kuboki, He, Kaneko, and Tokura}}]{Arima2005}
\bibinfo{author}{\bibfnamefont{T.}~\bibnamefont{Arima}},
  \bibinfo{author}{\bibfnamefont{J.~H.} \bibnamefont{Jung}},
  \bibinfo{author}{\bibfnamefont{M.}~\bibnamefont{Matsubara}},
  \bibinfo{author}{\bibfnamefont{M.}~\bibnamefont{Kuboki}},
  \bibinfo{author}{\bibfnamefont{J.~P.} \bibnamefont{He}},
  \bibinfo{author}{\bibfnamefont{Y.}~\bibnamefont{Kaneko}}, \bibnamefont{and}
  \bibinfo{author}{\bibfnamefont{Y.}~\bibnamefont{Tokura}},
  \bibinfo{journal}{J.\ Phys.\ Soc.\ Jpn.} \textbf{\bibinfo{volume}{74}},
  \bibinfo{pages}{1419} (\bibinfo{year}{2005}).

\bibitem[{\citenamefont{Matteo and Joly}(2006)}]{Matteo2006}
\bibinfo{author}{\bibfnamefont{S.~D.} \bibnamefont{Matteo}} \bibnamefont{and}
  \bibinfo{author}{\bibfnamefont{Y.}~\bibnamefont{Joly}},
  \bibinfo{journal}{Phys.\ Rev.\ B} \textbf{\bibinfo{volume}{74}},
  \bibinfo{pages}{014403} (\bibinfo{year}{2006}).

\bibitem[{\citenamefont{Igarashi and Nagao}(2010)}]{Igarashi2010}
\bibinfo{author}{\bibfnamefont{J.}~\bibnamefont{Igarashi}} \bibnamefont{and}
  \bibinfo{author}{\bibfnamefont{T.}~\bibnamefont{Nagao}},
  \bibinfo{journal}{J.\ Phys.\ Soc.\ Jpn.} \textbf{\bibinfo{volume}{79}},
  \bibinfo{pages}{014705} (\bibinfo{year}{2010}).

\end{thebibliography}

\end{document}